# Gamma Ray Bursts
# Maybe not so old after all

**Enrico Ramirez-Ruiz and William Lee**

**The discovery of a short-lived gamma-ray burst at a surprisingly early epoch in the history of the Universe shows how much is still unknown about the evolution of the parent systems of such bursts.**

Early on the morning of 14 July 2007, NASA's Swift spacecraft observed a brief flash of very high-energy photons, in an event known as a gamma-ay burst (GRB). The spacecraft's widefield gamma-ray monitor established the position of the burst, dubbed GRB 070714B, and within seconds had imaged it with its X-ray telescope, finding a rapidly fading source. Swift's prompt and highly accurate localization of the GRB allowed a tightly choreographed sequence of observations — performed using ground-based optical telescopes around the World — to follow the progressive dimming, or afterglow, of the source over the next day. These observations enabled Graham and colleagues, who report[1] their study of the GRB in *The Astrophysical Journal*, to firmly identify the galaxy in which the gamma-ray explosion took place: it is the most distant galaxy ever detected that harbours a GRB event of the short-lived kind.

Potentially distant (high-redshift) galaxies can be distinguished from their closer (lower redshift) counterparts through differences in their spectra. Ultraviolet light given off by hot, young stars in a far-off galaxy excites the surrounding gas clouds from which they were formed, producing strong emission lines in the galaxy's spectrum. The expansion of the Universe then shifts these lines to longer, redder wavelengths: by the time it reaches us, the wavelength of light from an object at redshift z has been stretched by a factor of 1+z. Has a spectral signature been unambiguously identified for the host galaxy of GRB 070714B, to claim a redshift record? Graham et al.[1] detected one emission line at optical wavelengths, and, on the basis of the colours of the host galaxy, deduced that it originated from ionized oxygen. But the redshift measurement is not of a quality that can be uncritically accepted as decisive. Traditionally, detection and identification of two spectral features is preferred. That said, Graham and colleagues' result seems quite robust.

Although each GRB is a unique event, these bursts fall roughly into one of two categories according to the duration of their emission at high energy: events lasting less than about two seconds are termed short, whereas the rest — the majority — are called long. GRB 070714B falls into the short-burst category, and being slightly more than 8 billion light years away, turns out to be the most distant explosion of this type found so far. Setting a new record as the farthest short GRB, together with a spectroscopic identification, might seem noteworthy enough, but it is not its distance alone that makes GRB 070714B interesting.

In astronomy, distance and time are inseparable, and because light travels at a finite speed, objects are seen as they were in the past. Distance, or equivalently, look-back time, is generally denoted by the redshift. GRB 070714B is detected at a redshift of z = 0.923, which means that it occurred when the Universe was about 40% of its present age, shortly after the time when most stars were being assembled in galaxies.

Swift was designed to answer the mystery of the short GRBs, which, despite being known for more than 40 years, have thus far not been linked to a specific progenitor, or associated with a particular type of galaxy that could be identified as a host. Therefore, constraining models for GRB formation through information about their environments has been challenging. By contrast, long GBRs have been rather firmly associated with the death of massive stars in relatively young galaxies[2,3].

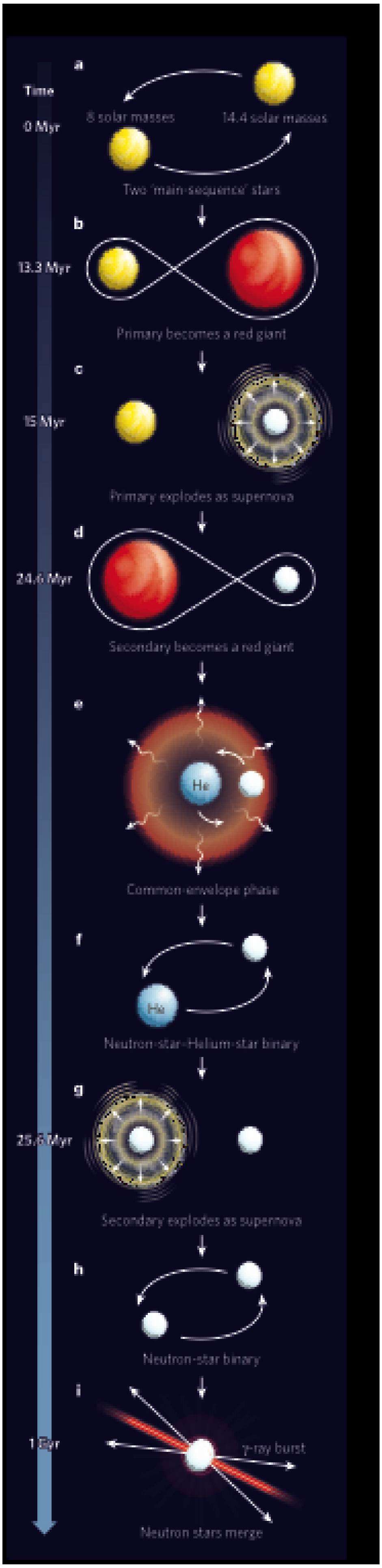

The most favoured models for short-GRB formation usually involved old systems, such as highly evolved binary-star progenitors containing neutron stars and/or black holes, which would go through one final cataclysm — hundreds to thousands of millions of years alter their formation — before quietly fading away[4]. And indeed, the first few short GRBs detected by Swift after its launch in 2004 seemed to be in general agreement with this picture: the host galaxies seemed to be old rather than young, devoid of much star formation, and generally in the declining phase of their cosmic evolution[5,6]. However, even at that early stage of Swift's observations, some diversity in host-galaxy type and cosmic epoch was already apparent[7], and because short bursts are intrinsically less luminous than their long counterparts, they may be prone to observational bias favouring detection at smaller distances, at which old systems are found.

In the years since Swift's first GRB observations, evidence has been steadily mounting for a more diverse landscape of how this class of explosion occurs[8]. Graham and colleagues' observation that GRB 070714B originated at an early epoch in the history of the Universe now strongly indicates that whereas the simple picture most often considered up to now — that short GRBs originate in old Systems — may account for a fraction of the events, it surely cannot explain them all. So what does this mean in terms of our understanding of the events themselves, their progenitors, and their hosts and environments?

**Figure 1. Possible evolutionary path to a short-lasting gamma-ray burst. a, b,** In a binary-star system consisting of two high-mass, 'main-sequence' stars, the more massive member (primary) exhausts its hydrogen nuclear-fuel supply, becomes a bloated red giant and transfers large amounts of mass to its lighter companion (secondary). **c,** The primary explodes as a supernova, leaving behind a neutron star. **d,** The secondary becomes a giant. **e,** Its expansion leads to a common-envelope phase in which the neutron star ploughs through the giant's outer layers, ejecting them from the system and leaving it in a tight orbit. **f,** The likely outcome is a neutron-star–helium-star binary. **g,** The helium star undergoes a supernova explosion. **h,** The resulting neutron-star pair is left in a short-period (about a day) orbit. **i,** With time, the system loses orbital energy and the stars merge, producing a gamma-ray burst; the time delay between the formation of the neutron-star binary and the burst occurrence is extremely sensitive to the separation of the stars after the common-envelope phase. The discovery[1] of GRB 070714B suggests that smaller separations than previously anticipated are needed. (Diagrams and temporal axis not to scale.)

For one thing, it clearly tells us that we have still much to learn about stellar evolution, particularly in binary systems, where stars live out their lives in very different ways from when they are travelling alone, like the Sun. In such systems, the time delay between progenitor formation and the onset of the burst, if one occurs, is extremely sensitive to the interaction between the stars during what is termed the common-envelope phase[9], when one star has become a bloated red giant and has essentially engulfed its ultradense, compact companion (Fig. 1). It has been suggested that this stage in the life of a binary system can lead to both short and long time delays[10], thus producing GRBs at both high and low redshift, respectively. Conversely, that the delays can be so varied as to be able to accommodate either progenitor population is indicative of how much remains to be discovered about binary stars. In this sense, identifying the progenitor of an event such as GRB 070714B would be a way of potentially probing different evolutionary pathways in the lives of stars.

Graham and colleagues' discovery also prompts us to consider other possible short-GRB formation mechanisms that may take place in both old and young galaxies. Among the possible candidates are highly magnetizad neutron stars called magnetars[11], events involving white dwarfs, and the interaction of compact stellar remnants in very densely populated environments such as stellar clusters. Whatever the mechanism, it is clear that singular events such as GRB 070714B present an exciting opportunity to study new regimes of stellar evolution at an epoch when galaxies were assembling most of their stars.

Over the coming years, space- and groundbased observations should allow us to uncover the detailed nature of these remarkable beasts. Astronomers no longer speak of short GRBs as utter mysteries. But that does not mean the puzzle is completely solved.

Enrico Ramirez-Ruiz is in the Department of Astronomy and Astrophysics, University of California, Santa Cruz, California 95064, USA.

William Lee is at the Instituto de Astronomía, Universidad Nacional Autónoma de México, México DF 04510, México.

e-mails:enrico@ucolick.org; wlee@astroscu.unam.mx